\documentclass[aps,prl,twocolumn,superscriptaddress,showpacs]{revtex4}

\usepackage{graphicx}
\usepackage{dcolumn}
\usepackage{bm}
\usepackage[dvips]{epsfig}
\usepackage{textcase}
\usepackage{xspace}
\newcommand\abbrev[1]{\textsc{\lowercase{#1}}\xspace}
\newcommand\PYTHIA{\abbrev{PYTHIA}}
\newcommand\ALPGEN{\abbrev{ALPGEN}}

\newcommand\GEANT{\abbrev{GEANT}}

\def \et {E_{T}}
\def  \met {\not\!\!\et }

\begin{document}


\title{

Search for $WW$ and $WZ$ resonances decaying to 
electron, missing $E_{T}$, and two jets in $p\overline{p}$
collisions at $\sqrt{s}=1.96$ TeV.

}

\affiliation{Institute of Physics, Academia Sinica, Taipei, Taiwan 11529, Republic of China} 
\affiliation{Argonne National Laboratory, Argonne, Illinois 60439, USA} 
\affiliation{University of Athens, 157 71 Athens, Greece} 
\affiliation{Institut de Fisica d'Altes Energies, Universitat Autonoma de Barcelona, E-08193, Bellaterra (Barcelona), Spain} 
\affiliation{Baylor University, Waco, Texas 76798, USA} 
\affiliation{Istituto Nazionale di Fisica Nucleare Bologna, $^{dd}$University of Bologna, I-40127 Bologna, Italy} 
\affiliation{Brandeis University, Waltham, Massachusetts 02254, USA} 
\affiliation{University of California, Davis, Davis, California 95616, USA} 
\affiliation{University of California, Los Angeles, Los Angeles, California 90024, USA} 
\affiliation{University of California, San Diego, La Jolla, California 92093, USA} 
\affiliation{University of California, Santa Barbara, Santa Barbara, California 93106, USA} 
\affiliation{Instituto de Fisica de Cantabria, CSIC-University of Cantabria, 39005 Santander, Spain} 
\affiliation{Carnegie Mellon University, Pittsburgh, Pennsylvania 15213, USA} 
\affiliation{Enrico Fermi Institute, University of Chicago, Chicago, Illinois 60637, USA}
\affiliation{Comenius University, 842 48 Bratislava, Slovakia; Institute of Experimental Physics, 040 01 Kosice, Slovakia} 
\affiliation{Joint Institute for Nuclear Research, RU-141980 Dubna, Russia} 
\affiliation{Duke University, Durham, North Carolina 27708, USA} 
\affiliation{Fermi National Accelerator Laboratory, Batavia, Illinois 60510, USA} 
\affiliation{University of Florida, Gainesville, Florida 32611, USA} 
\affiliation{Laboratori Nazionali di Frascati, Istituto Nazionale di Fisica Nucleare, I-00044 Frascati, Italy} 
\affiliation{University of Geneva, CH-1211 Geneva 4, Switzerland} 
\affiliation{Glasgow University, Glasgow G12 8QQ, United Kingdom} 
\affiliation{Harvard University, Cambridge, Massachusetts 02138, USA} 
\affiliation{Division of High Energy Physics, Department of Physics, University of Helsinki and Helsinki Institute of Physics, FIN-00014, Helsinki, Finland} 
\affiliation{University of Illinois, Urbana, Illinois 61801, USA} 
\affiliation{The Johns Hopkins University, Baltimore, Maryland 21218, USA} 
\affiliation{Institut f\"{u}r Experimentelle Kernphysik, Karlsruhe Institute of Technology, D-76131 Karlsruhe, Germany} 
\affiliation{Center for High Energy Physics: Kyungpook National University, Daegu 702-701, Korea; Seoul National University, Seoul 151-742, Korea; Sungkyunkwan University, Suwon 440-746, Korea; Korea Institute of Science and Technology Information, Daejeon 305-806, Korea; Chonnam National University, Gwangju 500-757, Korea; Chonbuk National University, Jeonju 561-756, Korea} 
\affiliation{Ernest Orlando Lawrence Berkeley National Laboratory, Berkeley, California 94720, USA} 
\affiliation{University of Liverpool, Liverpool L69 7ZE, United Kingdom} 
\affiliation{University College London, London WC1E 6BT, United Kingdom} 
\affiliation{Centro de Investigaciones Energeticas Medioambientales y Tecnologicas, E-28040 Madrid, Spain} 
\affiliation{Massachusetts Institute of Technology, Cambridge, Massachusetts 02139, USA} 
\affiliation{Institute of Particle Physics: McGill University, Montr\'{e}al, Qu\'{e}bec, Canada H3A~2T8; Simon Fraser University, Burnaby, British Columbia, Canada V5A~1S6; University of Toronto, Toronto, Ontario, Canada M5S~1A7; and TRIUMF, Vancouver, British Columbia, Canada V6T~2A3} 
\affiliation{University of Michigan, Ann Arbor, Michigan 48109, USA} 
\affiliation{Michigan State University, East Lansing, Michigan 48824, USA}
\affiliation{Institution for Theoretical and Experimental Physics, ITEP, Moscow 117259, Russia} 
\affiliation{University of New Mexico, Albuquerque, New Mexico 87131, USA} 
\affiliation{Northwestern University, Evanston, Illinois 60208, USA} 
\affiliation{The Ohio State University, Columbus, Ohio 43210, USA} 
\affiliation{Okayama University, Okayama 700-8530, Japan} 
\affiliation{Osaka City University, Osaka 588, Japan} 
\affiliation{University of Oxford, Oxford OX1 3RH, United Kingdom} 
\affiliation{Istituto Nazionale di Fisica Nucleare, Sezione di Padova-Trento, $^{ee}$University of Padova, I-35131 Padova, Italy} 
\affiliation{LPNHE, Universite Pierre et Marie Curie/IN2P3-CNRS, UMR7585, Paris, F-75252 France} 
\affiliation{University of Pennsylvania, Philadelphia, Pennsylvania 19104, USA}
\affiliation{Istituto Nazionale di Fisica Nucleare Pisa, $^{ff}$University of Pisa, $^{gg}$University of Siena and $^{hh}$Scuola Normale Superiore, I-56127 Pisa, Italy} 
\affiliation{University of Pittsburgh, Pittsburgh, Pennsylvania 15260, USA} 
\affiliation{Purdue University, West Lafayette, Indiana 47907, USA} 
\affiliation{University of Rochester, Rochester, New York 14627, USA} 
\affiliation{The Rockefeller University, New York, New York 10021, USA} 
\affiliation{Istituto Nazionale di Fisica Nucleare, Sezione di Roma 1, $^{ii}$Sapienza Universit\`{a} di Roma, I-00185 Roma, Italy} 

\affiliation{Rutgers University, Piscataway, New Jersey 08855, USA} 
\affiliation{Texas A\&M University, College Station, Texas 77843, USA} 
\affiliation{Istituto Nazionale di Fisica Nucleare Trieste/Udine, I-34100 Trieste, $^{jj}$University of Trieste/Udine, I-33100 Udine, Italy} 
\affiliation{University of Tsukuba, Tsukuba, Ibaraki 305, Japan} 
\affiliation{Tufts University, Medford, Massachusetts 02155, USA} 
\affiliation{Waseda University, Tokyo 169, Japan} 
\affiliation{Wayne State University, Detroit, Michigan 48201, USA} 
\affiliation{University of Wisconsin, Madison, Wisconsin 53706, USA} 
\affiliation{Yale University, New Haven, Connecticut 06520, USA} 
\author{T.~Aaltonen}
\affiliation{Division of High Energy Physics, Department of Physics, University of Helsinki and Helsinki Institute of Physics, FIN-00014, Helsinki, Finland}
\author{J.~Adelman}
\affiliation{Enrico Fermi Institute, University of Chicago, Chicago, Illinois 60637, USA}
\author{B.~\'{A}lvarez~Gonz\'{a}lez$^w$}
\affiliation{Instituto de Fisica de Cantabria, CSIC-University of Cantabria, 39005 Santander, Spain}
\author{S.~Amerio$^{ee}$}
\affiliation{Istituto Nazionale di Fisica Nucleare, Sezione di Padova-Trento, $^{ee}$University of Padova, I-35131 Padova, Italy} 

\author{D.~Amidei}
\affiliation{University of Michigan, Ann Arbor, Michigan 48109, USA}
\author{A.~Anastassov}
\affiliation{Northwestern University, Evanston, Illinois 60208, USA}
\author{A.~Annovi}
\affiliation{Laboratori Nazionali di Frascati, Istituto Nazionale di Fisica Nucleare, I-00044 Frascati, Italy}
\author{J.~Antos}
\affiliation{Comenius University, 842 48 Bratislava, Slovakia; Institute of Experimental Physics, 040 01 Kosice, Slovakia}
\author{G.~Apollinari}
\affiliation{Fermi National Accelerator Laboratory, Batavia, Illinois 60510, USA}
\author{J.~Appel}
\affiliation{Fermi National Accelerator Laboratory, Batavia, Illinois 60510, USA}
% ...

\collaboration{CDF Collaboration\footnote{With visitors from $^a$University of Massachusetts Amherst, Amherst, Massachusetts 01003,
$^b$Universiteit Antwerpen, B-2610 Antwerp, Belgium, 
$^c$University of Bristol, Bristol BS8 1TL, United Kingdom,
$^d$Chinese Academy of Sciences, Beijing 100864, China, 
$^e$Istituto Nazionale di Fisica Nucleare, Sezione di Cagliari, 09042 Monserrato (Cagliari), Italy,
$^f$University of California Irvine, Irvine, CA  92697, 
$^g$University of California Santa Cruz, Santa Cruz, CA  95064, 
$^h$Cornell University, Ithaca, NY  14853, 
$^i$University of Cyprus, Nicosia CY-1678, Cyprus, 
$^j$University College Dublin, Dublin 4, Ireland,
$^k$University of Edinburgh, Edinburgh EH9 3JZ, United Kingdom, 
$^l$University of Fukui, Fukui City, Fukui Prefecture, Japan 910-0017,
$^m$Kinki University, Higashi-Osaka City, Japan 577-8502,
$^n$Universidad Iberoamericana, Mexico D.F., Mexico,
$^o$University of Iowa, Iowa City, IA  52242,
$^p$Kansas State University, Manhattan, KS 66506,
$^q$Queen Mary, University of London, London, E1 4NS, England,
$^r$University of Manchester, Manchester M13 9PL, England,
$^s$Muons, Inc., Batavia, IL 60510, 
$^t$Nagasaki Institute of Applied Science, Nagasaki, Japan, 
$^u$University of Notre Dame, Notre Dame, IN 46556,
$^v$Obninsk State University, Obninsk, Russia,
$^w$University de Oviedo, E-33007 Oviedo, Spain, 
$^x$Texas Tech University, Lubbock, TX  79609, 
$^y$IFIC(CSIC-Universitat de Valencia), 56071 Valencia, Spain,
$^z$Universidad Tecnica Federico Santa Maria, 110v Valparaiso, Chile,
$^{aa}$University of Virginia, Charlottesville, VA  22906,
$^{bb}$Bergische Universit\"at Wuppertal, 42097 Wuppertal, Germany,
$^{cc}$Yarmouk University, Irbid 211-63, Jordan,
$^{kk}$On leave from J.~Stefan Institute, Ljubljana, Slovenia, 
}}
\noaffiliation



\begin{abstract}

Using data from 2.9~fb$^{-1}$ of integrated luminosity collected with
the CDF~II detector at the Tevatron, we search for resonances decaying
into a pair of on-shell gauge bosons, $WW$ or $WZ$, where one $W$
decays into an electron and a neutrino, and the other boson decays
into two jets.  We observed no statistically significant excess above
the expected standard model background, and we set cross section
limits at 95\% confidence level on $G^{*}$ (Randall-Sundrum graviton),
$Z'$, and $W'$ bosons. By comparing these limits to theoretical cross
sections, mass exclusion regions for the three particles are
derived. The mass exclusion regions for $Z'$ and $W'$ are further
evaluated as a function of their gauge coupling strength.

\end{abstract}

\pacs{14.70.Pw, 13.85.Rm, 13.85.Qk, 14.80.-j}

\maketitle

\newpage
Some models of new physics beyond the standard model predict particles
that decay into pairs of on-shell bosons, for example $Z'$,
$W'$\cite{c:etgauge}, or the Randall-Sundrum graviton
$G^{*}$~\cite{c:graviton}. Searches for these particles in different
decay channels have been reported
elsewhere~\cite{c:lep,c:gk,c:zk,c:bk,c:wd0}. Most of them used final states
consisting of only leptons or photons. In this Letter we search for
these particles in the form of diboson resonances where one boson is a
$W$ decaying into an electron and a neutrino, and the other is a $W$ or
$Z$ which decays into two jets. This search has the advantage of
detecting two types of diboson resonances, $WW$ and $WZ$, with the
same final-state topology. The hadronic decay mode of the $W$ or $Z$ to two
jets has a higher branching fraction compared to the leptonic mode;
however, the background from jets also increases. Thus we implement a
selection based on transverse energy ($E_{T}$)~\cite{c:cdf_coord} of
the detected objects in the final state to reduce standard model
backgrounds and enhance sensitivity.

The diboson decay modes of $Z'$ and $W'$ directly probe the gauge
coupling strength between the new and the standard model gauge
bosons. The coupling strength strongly influences the decay branching
ratios and the natural widths of the new gauge bosons. In an extended
gauge model theory~\cite{c:etgauge} the standard model coupling
strength, $g\cdot\cos\theta_{w}$, is replaced by $\xi\cdot
g\cdot\cos\theta_{w}$, where $\xi = C\cdot(M_{W}/M_{V})^{2}$; $C$ is a
parameter that sets the coupling strength; and $M_{V}$ is the mass of
the new gauge boson, $Z'$ or $W'$.  We set cross section limits on
$Z'$ and $W'$ as a function of mass and of $\xi$. Our results extend
the sensitivity beyond the CDF Run I $W'$ results~\cite{c:wprime} with
almost 30 times the integrated luminosity, and, for the first time,
set $Z'$ limits as a function of mass and gauge coupling strength.
For $G^{*}$, the coupling constant $k/\overline{M}_{Pl}$ dictates the
branching ratio and natural width~\cite{c:graviton}, where $k$ and
$\overline{M}_{Pl}$ are respectively the curvature of the extra
dimension and the reduced Planck mass scale. This is also the first
search for the $G^{*}$ in the $WW$ decay mode.

This analysis is based on data corresponding to an integrated
luminosity of 2.9~fb$^{-1}$ collected using the CDF~II detector
between March 2002 and February 2008. The detector is approximately
forward-backward and azimuthally symmetric. The detector elements
relevant to this analysis are the tracking system and the
calorimeters. The tracking system consists of an eight-layer silicon
tracker~\cite{c:silicon} surrounded by a 96-layer open-cell drift
chamber (COT)~\cite{c:cot}. The fiducial coverage of the COT is
$|\eta|<1.0$~\cite{c:cdf_coord}, and the silicon detector extends the
coverage to $|\eta|<2.0$. The integrated tracking system is contained
within a superconducting solenoid, providing a 1.4 T magnetic
field. Surrounding the tracking system are the electromagnetic (EM)
and hadronic calorimeters~\cite{c:calo}, divided into ``central''
($|\eta|<1.1$) and ``plug'' ($1.1<|\eta|<3.6$) regions. The
calorimeters are made of lead (EM) and iron (hadronic) absorbers
sandwiched between plastic scintillators that provide measurements of
shower energies. At approximately the shower maximum, the EM
calorimeters contain fine-grained detectors~\cite{c:shower_max} for
measuring shower positions and profiles.


As we are looking for events with an electron, a neutrino, and two
jets, we start with data that were collected with an online selection
requirement of a central electron with $|\eta|<1$ and
$E_{T}>18$~GeV. From this dataset we select events that have an
isolated electron~\cite{c:iso} with $E_{T}>30$ GeV, a neutrino
identified by the requirement that the missing $E_{T} (\met) >30$ GeV,
two or three jets with $|\eta|<2.5$ and $E_{T}>30$ GeV, and an overall
$H_{T}>150$ GeV, where $H_{T}$ is the scalar sum of the electron
$E_{T}$, the $\met$, and the $E_{T}$ of all jets~\cite{c:ht}.

To form a $WW$ or $WZ$ hypothesis for the selected events, the
electron and $\met$ are first combined to form a $W$
candidate. Because the longitudinal component of the neutrino momentum
($E_{z}^{\nu}$) is not available, the invariant mass of the electron
and $\met$ is artificially set to the $W$ mass. With this assumption,
the conservation of energy and momentum results in a quadratic
equation for $E_{z}^{\nu}$. If the discriminant of the quadratic
equation is negative, the combination is discarded. If it is positive,
there are two solutions and both are kept. In addition, two jets are
combined to form a second $W$ candidate or a $Z$ candidate. In the
case of a $W$ candidate, we require the two-jet invariant mass
($M_{jj}$) to fall between 65 and 95 GeV$/c^{2}$, corresponding to
$\pm 1.5 \sigma$ of the expected reconstructed $W$ resolution. In the
case of a $Z$ candidate, this window is between 75 and 105
GeV$/c^{2}$.  For a three-jet event, there are three two-jet invariant
mass combinations. In this case only the pair with the invariant mass
closest to either the $W$ or the $Z$ mass is kept in order to reduce
the combinatorial background.  The reconstructed $W$ or $Z$ candidates
are then combined to form the final $WW$ or $WZ$ invariant mass.

Twelve standard model processes are considered as background for this
analysis: $W$($\rightarrow e^{\pm}\nu$)+jets, QCD jets, $t\overline{t}$,
$WW$, $Z$($\rightarrow e^{+}e^{-}$)+jets,
$W$($\rightarrow\tau^{\pm}\nu$)+jets, single top, $WZ$, $W\gamma$,
$Z\rightarrow\tau^{+}\tau^{-}$, $\gamma\gamma$, and $ZZ$.  The
dominating background is $W$+jets whose contribution is estimated by
Monte Carlo simulation using the \ALPGEN~\cite{c:alpgen} event
generator, interfaced to \PYTHIA~\cite{c:pythia} for parton showering
and followed by the \GEANT~3~\cite{c:geant} based CDF~II detector
simulation. With the exception of the QCD jet background, the rest of
the background processes are all estimated by Monte Carlo simulation using
the \PYTHIA event generator. The cross sections used for the simulated
background processes are obtained from NLO calculations.

\begin{figure}
{\psfig{file=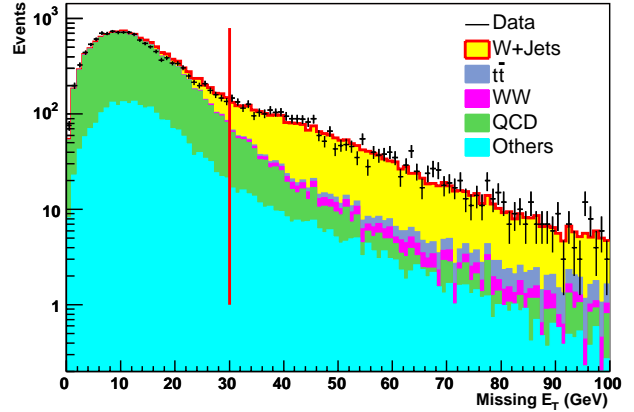,width=8.5cm}}
\caption{$\met$ spectrum from events with two jets. The vertical line
  marks the $\met>30$ GeV cut. The QCD component is scaled such that
  data and expected background match at the peak area where no signal
  is expected. ``Others'' background includes: $Z$($\rightarrow
  e^{+}e^{-}$)+jets, single top, $WZ$, $W\gamma$,
  $Z\rightarrow\tau^{+}\tau^{-}$, $\gamma\gamma$, and $ZZ$.}
\label{f:met}
\end{figure}

The QCD jet background comes from events with three or more jets where
one of the jets is misidentified as an electron. With this misidentified
electron, the event may pass through subsequent event selection
criteria and the reconstruction processes. The contribution of the QCD
jet background is estimated using a dataset that has an online
selection requirement of one jet with $E_{T} > 20$ GeV. We first exclude
events that have any identified electrons, then each jet in the
central region is treated as an electron with a weight corresponding
to the probability that a jet is misidentified as an electron. This
probability is a function of jet $E_{T}$ and varies from $~10^{-4}$ at
30 GeV to $~10^{-3}$ above 100 GeV~\cite{c:bk}. The misidentified
electron is combined with the $\met$ and then with two jets to form $WW$
or $WZ$ candidates as described earlier. The resulting QCD jet
background is normalized to the data by matching the $\met$ spectrum
between data and expected background at their peaks around 10 GeV,
where little signal is expected and the QCD jet background
dominates. This normalization factor is used for the QCD jet
contribution throughout the analysis. Figure~\ref{f:met} shows the
resulting $\met$ spectrum for events with two jets that would have
passed the event selection criteria except for the $\met>30$ GeV cut.

The systematic uncertainties taken into account in the background
calculations are the following, listed by decreasing significance: jet
energy scale (JES) uncertainty~\cite{c:jes}, theoretical cross section
uncertainty~\cite{c:thx}, luminosity uncertainty~\cite{c:lum}, and jet
misidentification rate uncertainty.  The dominating systematic
uncertainty is the JES uncertainty which amounts to $\sim13\%$ of the
estimated background. The cross section and luminosity uncertainties are
$\sim6\%$ each.

Signal detection efficiencies are also determined from simulated
events using the \PYTHIA event generator. For a set of selected mass
values ranging from 165 GeV$/c^{2}$ to 1000 GeV$/c^{2}$, the three
types of particles are simulated: $G^{*}$ with
$k/\overline{M}_{Pl}=0.1$, $Z'$ and $W'$ with \PYTHIA default settings
corresponding to the extended gauge model with a suppression factor
$\xi = (M_{W}/M_{V})^{2}$, i.e., $C=1$.  The reconstructed signals are
Gaussian in shape, and the mass resolution is linearly proportional to
the generated mass values, varying from 20 GeV$/c^{2}$ at 200
GeV$/c^{2}$ mass to 80 GeV$/c^{2}$ at 1000 GeV$/c^{2}$ mass. For
calculating the efficiencies we choose an acceptance mass window
corresponding to $\pm1.5$ times the reconstructed signal
resolution. This choice gives a good signal to background ratio. The
same acceptance mass windows are also used to obtain the number of
background events.

\begin{figure*}[h]
{\psfig{file=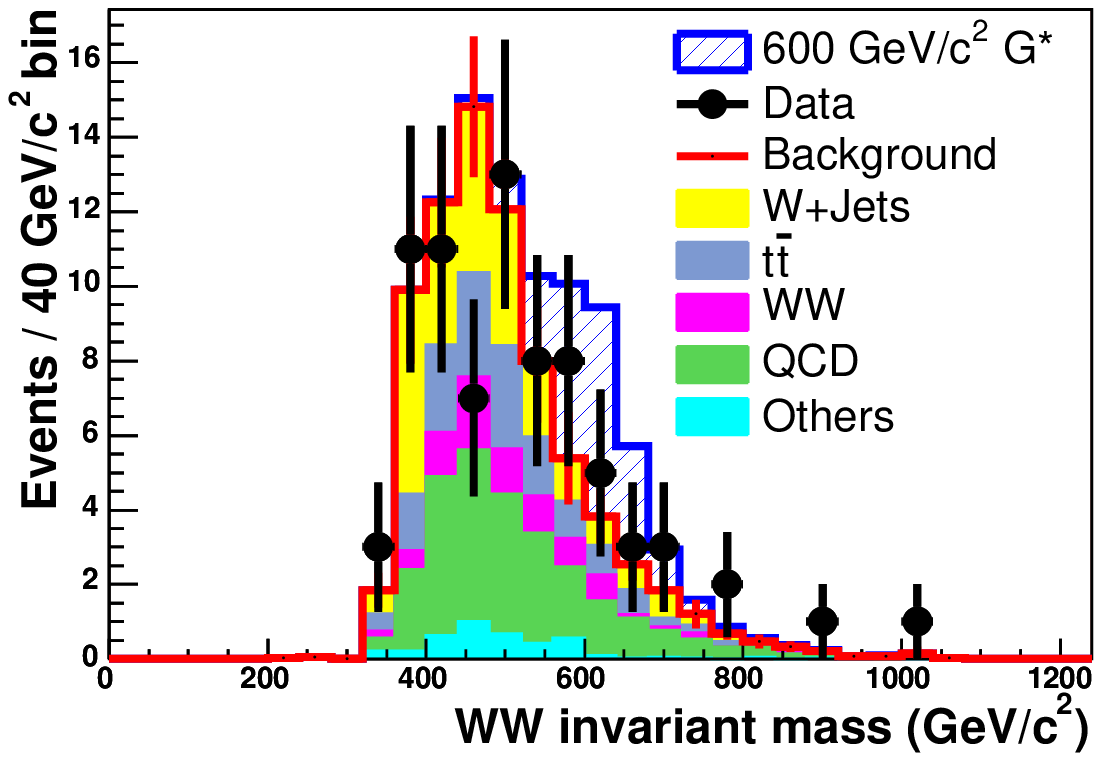,width=5.7cm,}}
{\psfig{file=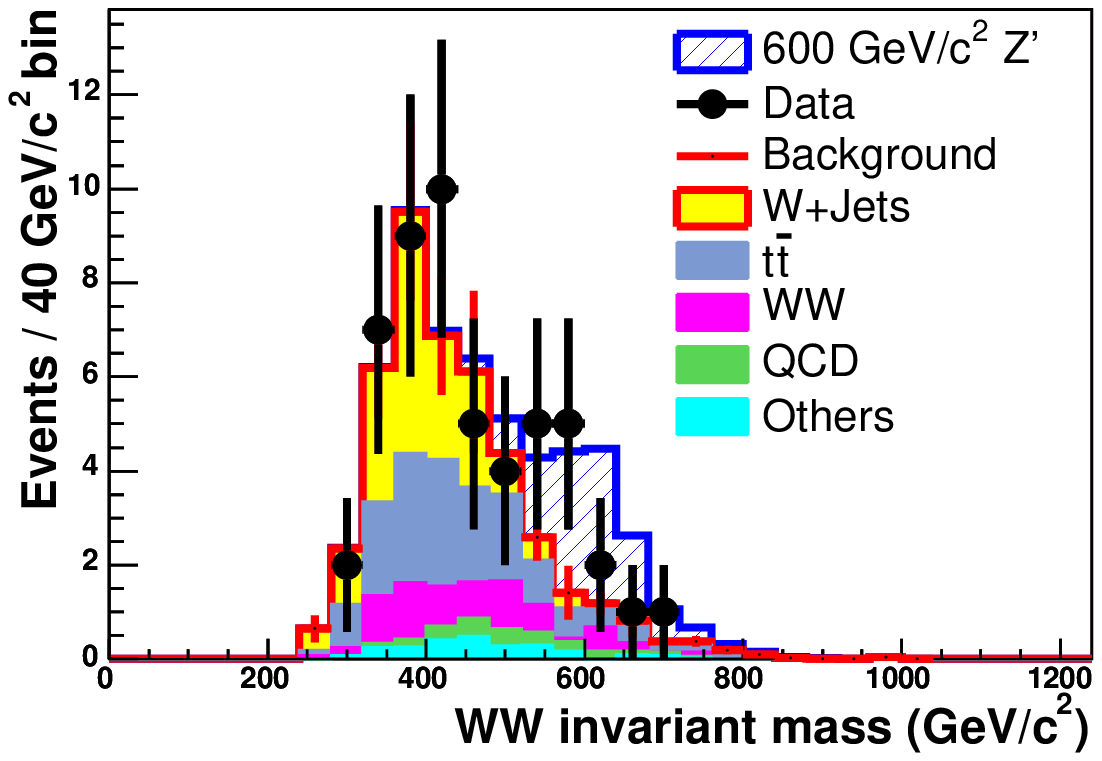,width=5.7cm,}}
{\psfig{file=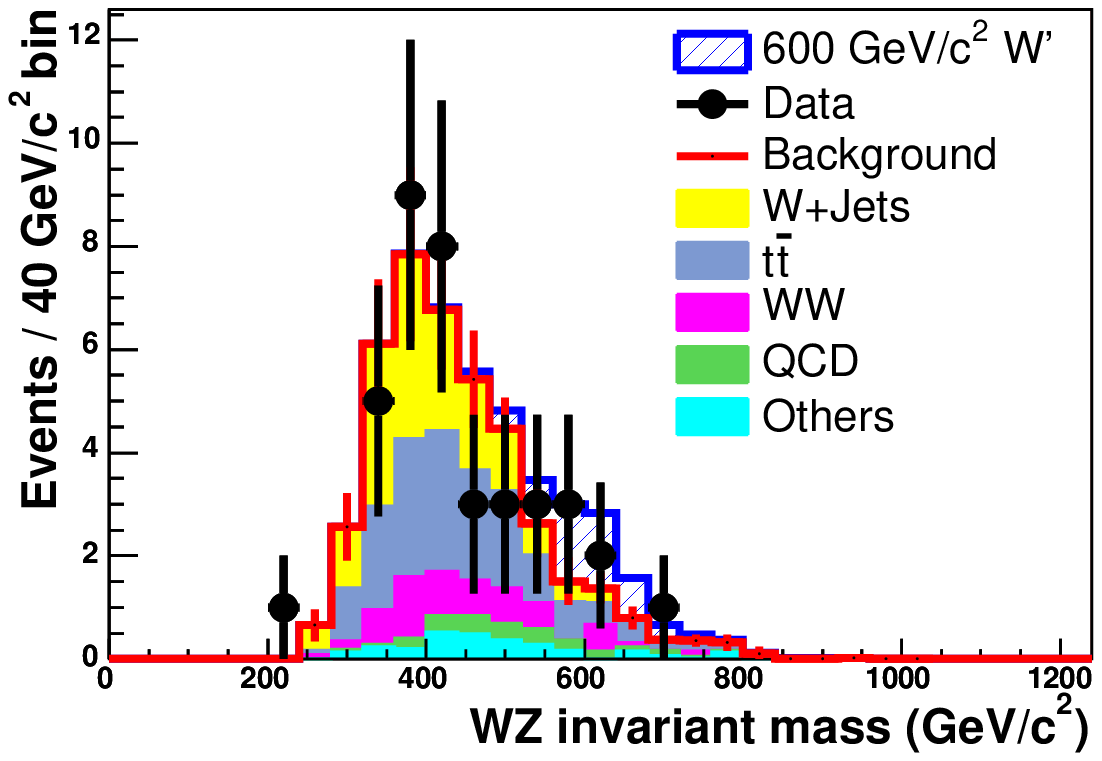,width=5.7cm,}}
\caption{ Invariant mass distributions with optimal set of $E_{T}$
  cuts for 600 GeV$/c^{2}$ signals.  Left to right: $G^{*}(WW)$,
  $Z'(WW)$, $W'(WZ)$.  $G^{*}$ and
  $Z'$ are the same decay mode ($WW$) but with
  different optimal selections. $Z'$ and $W'$ have the same
  optimal selection but different decay modes ($WW$ vs
  $WZ$). Superimposed on top of the backgrounds are
  signals at 600 GeV$/c^{2}$ mass corresponding to the theoretical cross
  sections.}
\label{f:g}
\end{figure*}
\begin{figure*}[h]
{\psfig{file=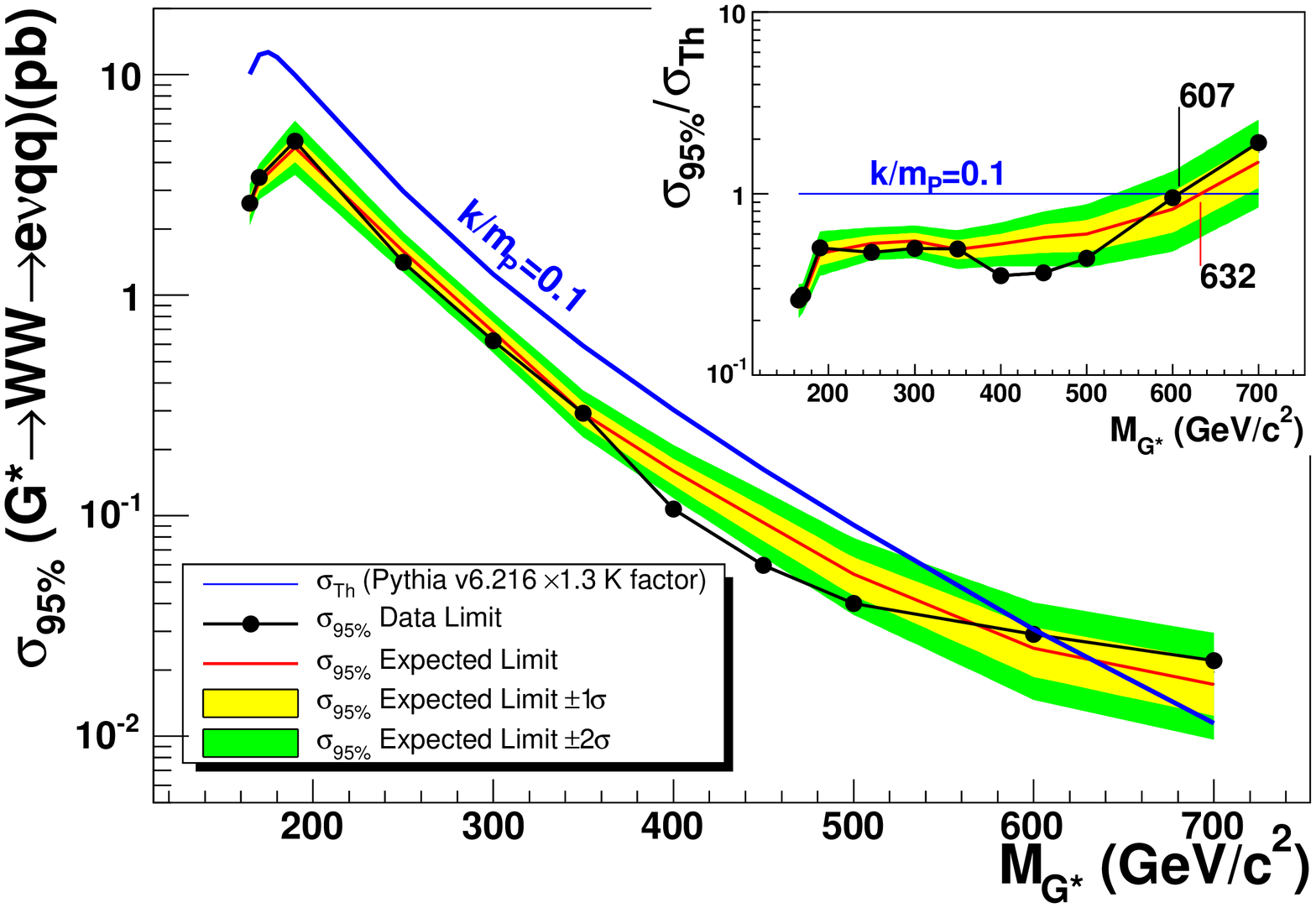,width=5.7cm}}
{\psfig{file=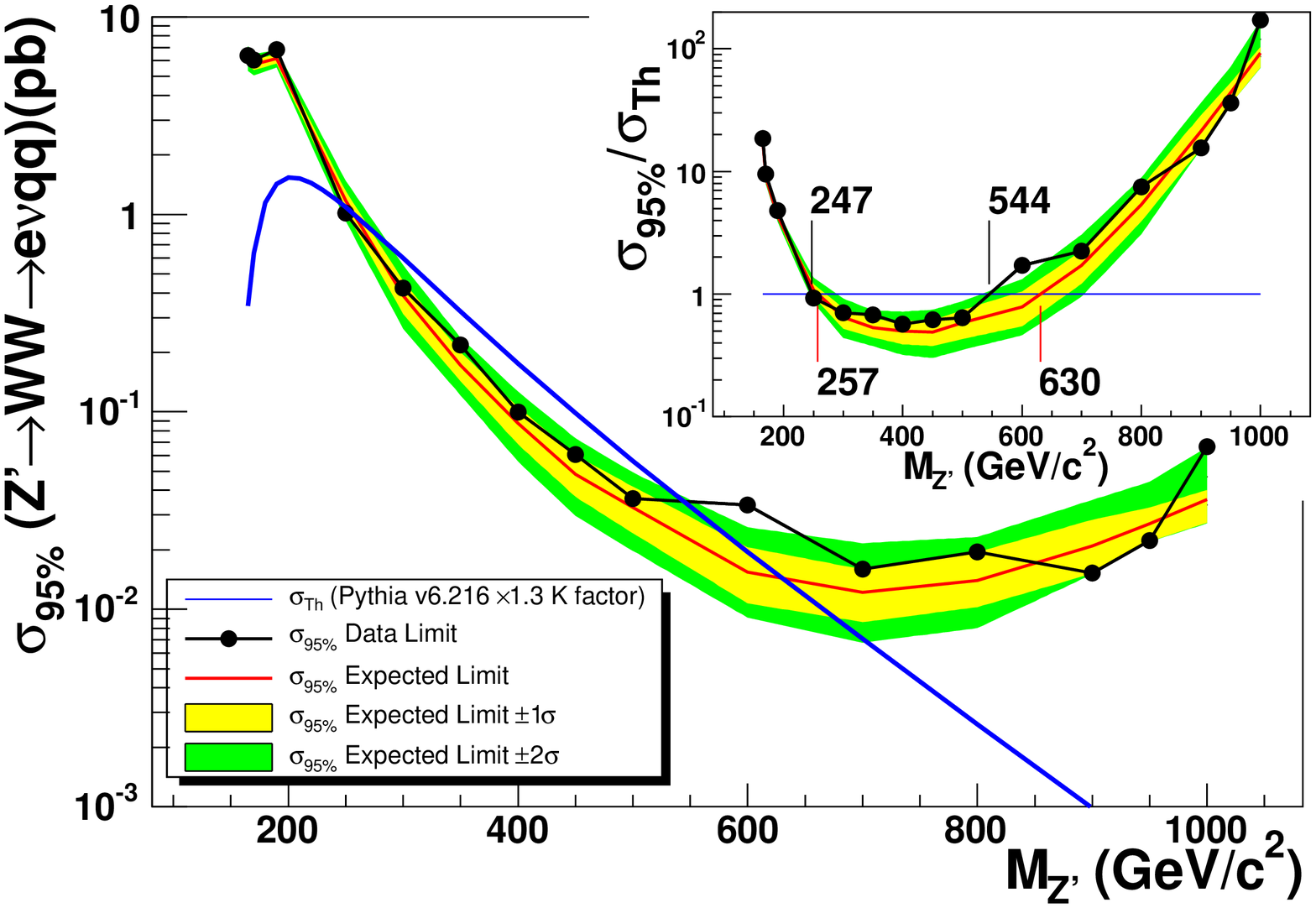,width=5.7cm}}
{\psfig{file=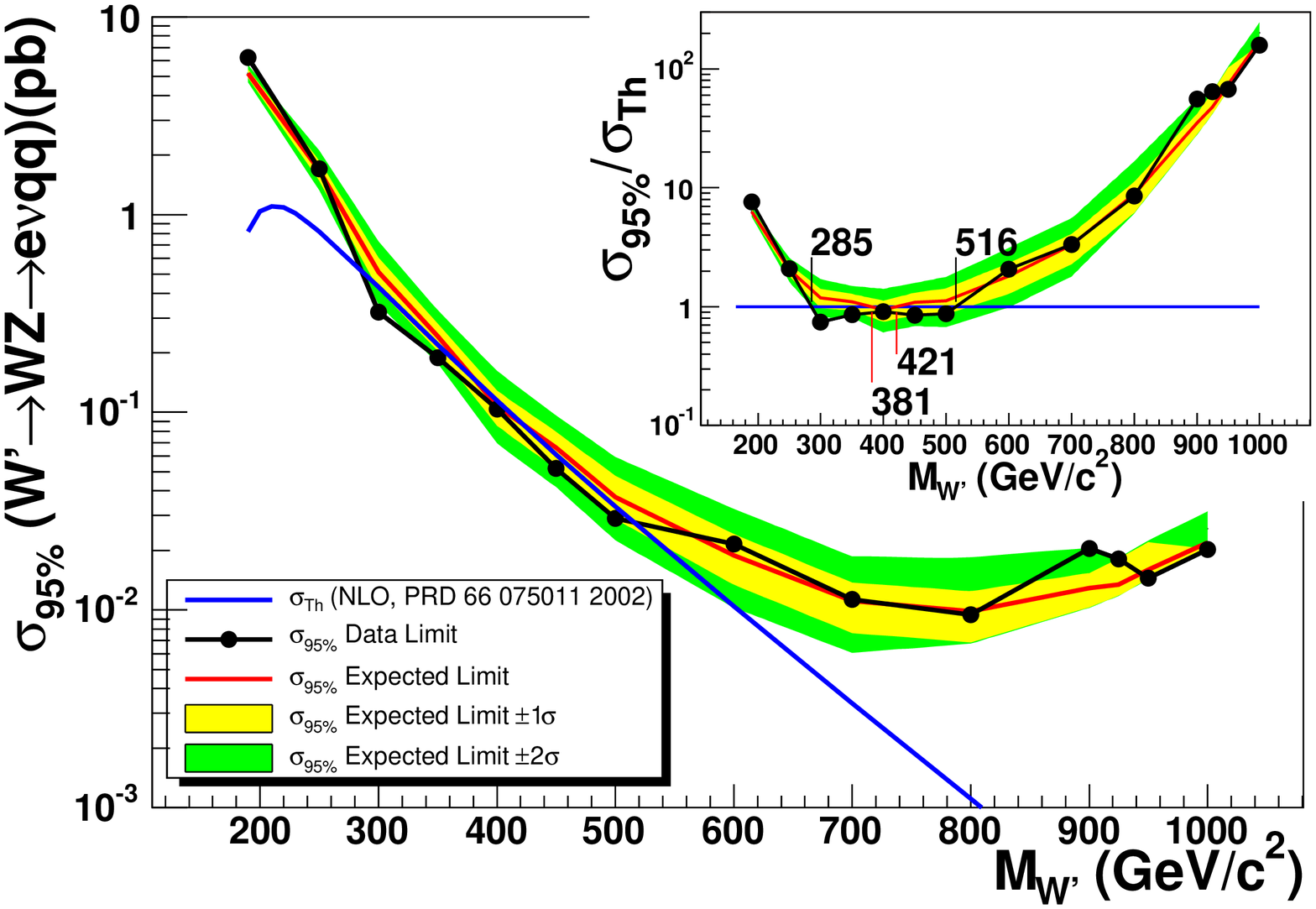,width=5.7cm}}
\caption{ Cross section limits at 95\% C.L. Left to right: $G^{*}$,
  $Z'$, $W'$. Inserts at upper right are cross section limits divided
  by the theoretical cross sections. Yellow and green bands show
  $1\sigma$ and $2\sigma$ bands on the expected limits, respectively.}
\label{f:l}
\end{figure*}
\begin{figure*}[!hp]
{\psfig{file=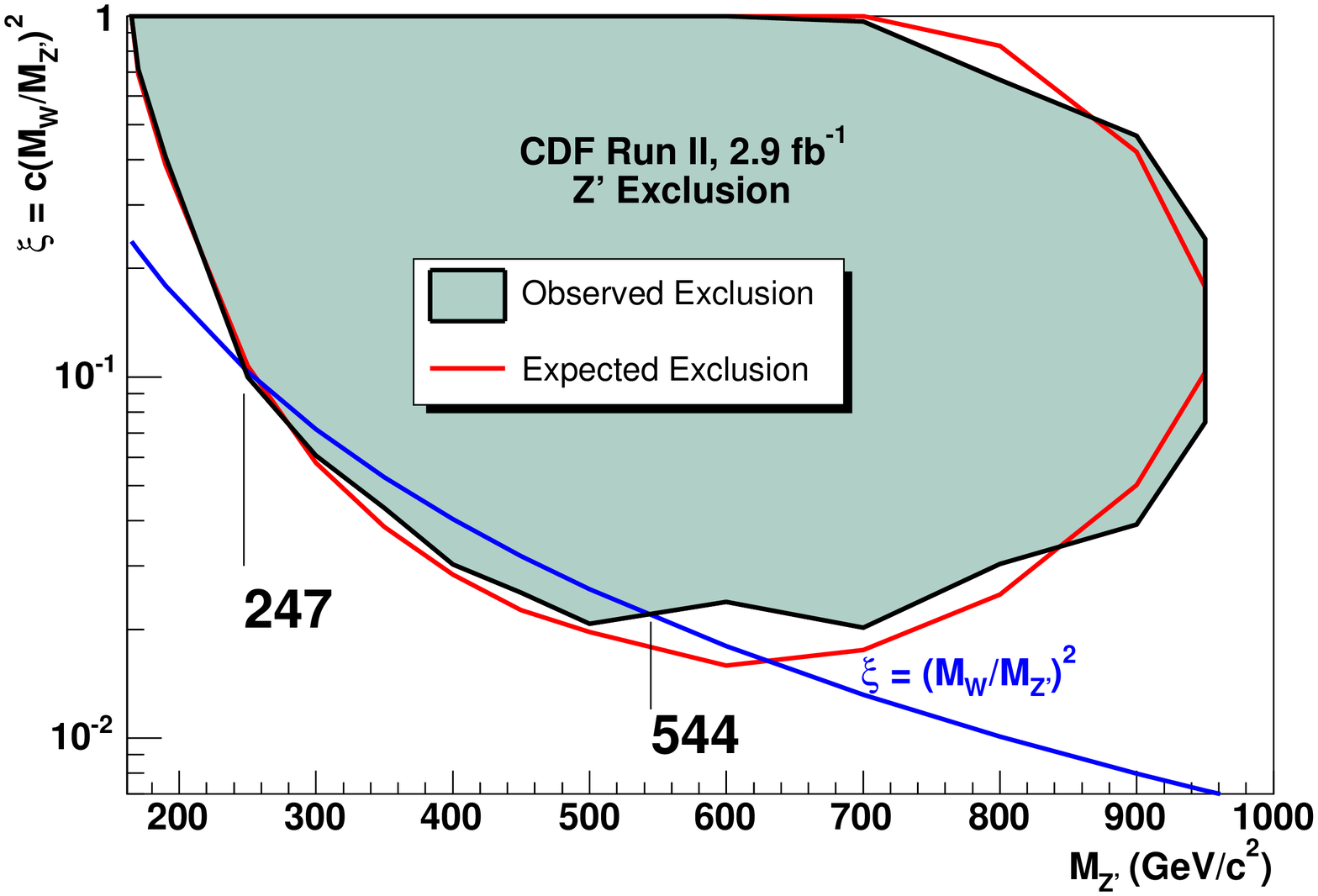,width=8.0cm}}
{\psfig{file=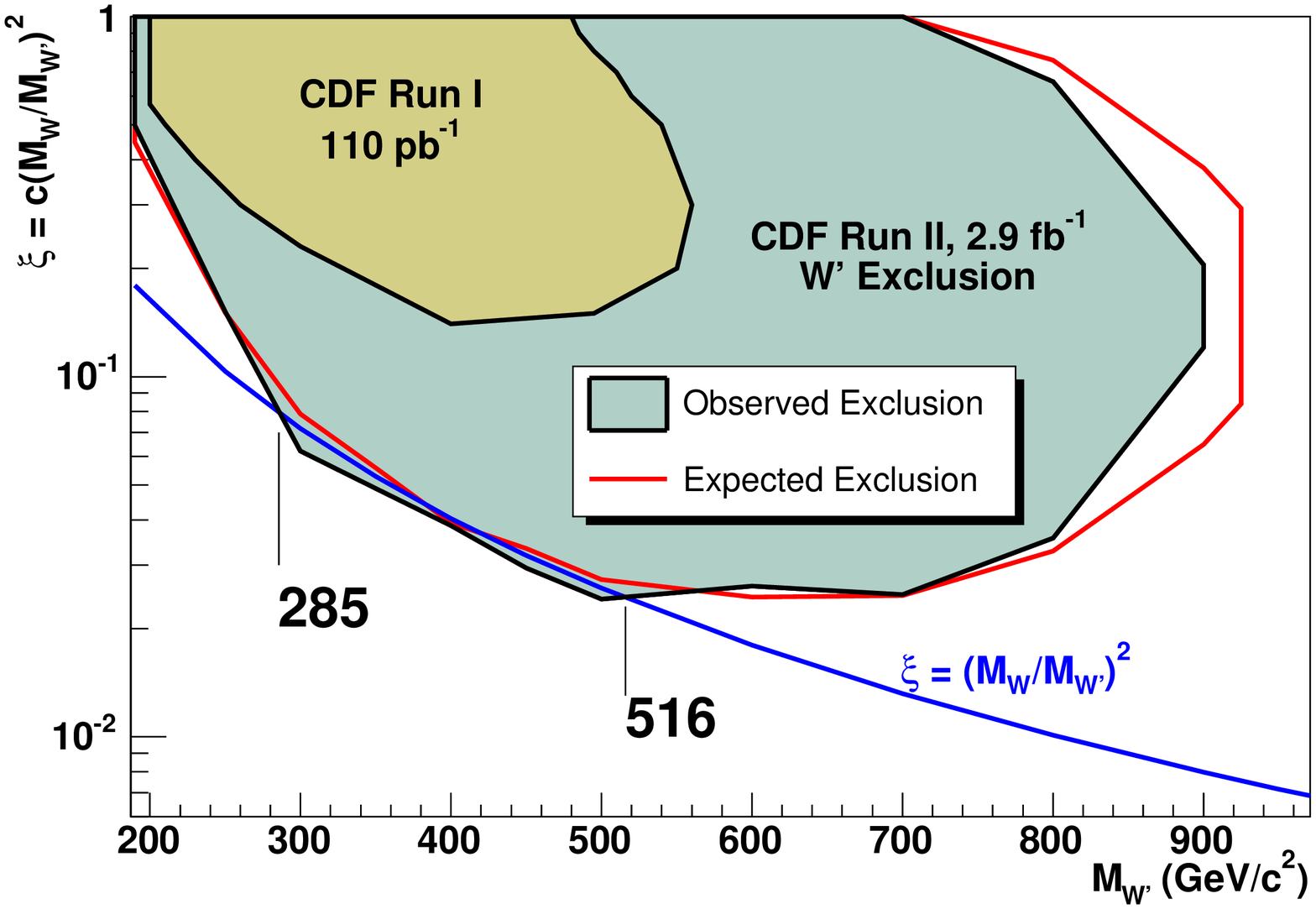,width=8.0cm}}
\caption{ $Z'$ (left) and $W'$ (right) exclusion regions
  as a function of mass and $\xi$. The $\xi = (M_{W}/M_{V})^{2}$ (i.e., $C=1$) lines
  indicate \PYTHIA defaults and are commonly used for mass exclusion
  regions. The vertical lines mark the results as shown in
  Table~\ref{t.mex}. Also shown in the $W'$ plot is the CDF Run
  I result.}
\label{f:c}
\end{figure*}

The systematic uncertainties taken into account for the signal
acceptance, defined as the product of signal detection efficiency and
integrated luminosity, in order of decreasing significance, are: jet
energy scale (JES) uncertainty, luminosity uncertainty, initial state
radiation (ISR) uncertainty, final state radiation (FSR) uncertainty,
and parton distribution function (PDF) uncertainty.  Similarly to the
background uncertainties, the JES uncertainty dominates the systematic
uncertainties and varies from $~12$\% at 170 GeV$/c^{2}$ mass to
$~6$\% at 700 GeV$/c^{2}$ mass for $G^{*}$, $~13$\% (170 GeV$/c^{2}$)
to $~6$\% (1000 GeV$/c^{2}$) for $Z'$, and $~9$\% (190 GeV$/c^{2}$) to
$~6$\% (1000 GeV$/c^{2}$) for $W'$. ISR, FSR and PDF uncertainties are
of the order of $~1-3$\% each and decrease with increasing diboson
mass.

In order to improve sensitivity at higher mass, additional sets of
higher $E_{T}$ cuts for the constituent particles (observed in the
detector as electron, $\met$ from neutrino, and jets) are tried. Two
series of the $E_{T}$ cut sets are implemented. The first series
requires a higher $E_{T}$ on all four participating particles ranging
from 40 GeV to 80 GeV in steps of 10 GeV. The second series requires a
higher $E_{T}$ on only one daughter particle from each of the decaying
bosons, i.e., a higher $E_{T}$ for either the electron or the
neutrino, and the same higher $E_{T}$ for one of the two jets. The
$E_{T}$ values in this series range from 40 GeV to 120 GeV in steps of
10 GeV. For each set of $E_{T}$ cuts the systematic uncertainties for
the backgrounds and the acceptances are re-evaluated, but are found to
be not very sensitive to the variations.

To find the optimal set of $E_{T}$ cuts at each selected mass point,
the expected cross section limits, which are based only on the
background and the signal acceptance, are calculated for each set of
cuts. We found that the first series of $E_{T}$ cuts gives the best
expected limits for $Z'$ and $W'$, while the second series is best for
$G^{*}$. The optimal $E_{T}$ cuts for each particle type are then
selected from their own optimal series. The sets that give the best
expected limits are chosen without reference to their impact on the
data sample. Although the background processes respond differently to
the two series of $E_{T}$ cuts, the best expected limits obtained from
each series are very similar. Generally, as the mass increases the
higher $E_{T}$ cuts yield better expected limits.

We use a Bayesian method~\cite{c:genlimit} to calculate cross section
limits. Inputs to the calculation are signal acceptance, estimated
background, and observed data. The signal acceptance and background
are assigned priors and modeled via a Monte Carlo method that allows
correlation of uncertainties between acceptance and background. In our
analysis, the JES and luminosity uncertainties in the acceptance and
in the background are correlated. The expected limits are calculated
by simulating observed data based on the expected background with
Poisson fluctuations.

Figure~\ref{f:g} shows typical invariant mass distributions
reconstructed for each particle type for a mass of 600 GeV$/c^{2}$
using the optimal set of $E_{T}$ cuts in each case. The $WW$ invariant
mass distributions are shown for $G^{*}$ and $Z'$, and the $WZ$
invariant mass distribution is shown for $W'$. The optimal set of
$E_{T}$ cuts for $G^{*}$ at 600 GeV$/c^{2}$ is from the second series
with $E_{T} > 120$ GeV, while both $Z'$ and $W'$ favor the first
series with $E_{T} > 60$ GeV. The background compositions, as shown in
Table~\ref{t.bg}, are found to be more sensitive to the different sets
of $E_{T}$ cuts than to the different decay types ($WW$ or $WZ$). For
$Z'$ and $W'$, the QCD jet background has a much lower contribution
owing to the stricter $E_{T}$ requirements.

\begin{table}[h]
\begin{center}
\caption{\label{t.bg} Percentage fractional background compositions
  in Fig.~\ref{f:g}. The uncertainties include both statistical and
  systematic uncertainties. }
\begin{tabular}{ccrclcrclcrcl}\hline\hline
                 && \multicolumn{3}{c}{$G^{*}(WW)$} &&  \multicolumn{3}{c}{$Z'(WW)$} && \multicolumn{3}{c}{$W'(WZ)$} \\ \hline
$W$+jets         && 31.8&$\pm$&8.2  && 33.0&$\pm$&10.0  && 36.8&$\pm$&9.7  \\ 
$t\overline{t}$  && 19.6&$\pm$&2.7  && 35.1&$\pm$&4.0   && 37.4&$\pm$&5.2  \\ 
$WW$             && 10.7&$\pm$&3.2  && 15.2&$\pm$&2.8   && 13.4&$\pm$&3.2  \\
QCD jets         && 32.7&$\pm$&6.5  && 5.1&$\pm$&1.0    && 5.6&$\pm$&1.1   \\
Others           && 5.3&$\pm$&0.9   && 2.4&$\pm$&0.9    && 3.4&$\pm$&1.0   \\
\hline \hline                                                       
\end{tabular}
\end{center}
\end{table}    

Without a statistically significant excess above the expected
background in the invariant mass plots, we calculate the cross section
limits at 95\% confidence level (C.L.) for the observed data.
Figure~\ref{f:l} shows the observed and the expected 95\% C.L. cross
section limits overlaid with theoretical cross sections.  The
theoretical cross sections for $G^{*}$ and $Z'$ are calculated from
\PYTHIA version 6.216, and a constant K-factor of 1.3 is applied to
take into account the NLO correction~\cite{c:gk,c:zk,c:bk}. The
theoretical cross section for $W'$ is derived from a NLO
calculation~\cite{c:wk}. The upper right inserts in Fig.~\ref{f:l}
show ratios of the limits to the theoretical cross sections. Where the
ratio is below one the mass region is excluded. Table~\ref{t.mex}
summarizes the mass exclusion regions from the figures.

\begin{table}[h]
\begin{center}
\caption{\label{t.mex} Mass exclusion region at 95\% C.L. with $k/\overline{M}_{Pl}=0.1$ for
  $G^{*}$, and $\xi = (M_{W}/M_{V})^{2}$ ($C=1$) for $Z'$ and $W'$.}
\begin{tabular}{ccccccc}\hline\hline
                                  && $G^{*}$ && $Z'$     && $W'$ \\ \hline
Expected Exclusion (GeV$/c^{2}$)  && $<632$  && 257-630  && 381-421  \\ 
Observed Exclusion (GeV$/c^{2}$)  && $<607$  && 247-544  && 285-516  \\ 
\hline \hline 
\end{tabular}
\end{center}
\end{table}    

The results shown in Fig.~\ref{f:l} and Table~\ref{t.mex} for $Z'$ and
$W'$ are based on a gauge coupling mixing factor of $\xi =
C\cdot(M_{W}/M_{V})^{2}$, with $C=1$. Since signal acceptance is the
only quantity that changes with $\xi$ in the cross section limit
calculation, at each mass point we re-evaluate signal acceptances for
different $\xi$ values and calculate cross section limits as a
function of $\xi$. Comparing the calculated and theoretical cross
sections as a function of $\xi$, a $\xi$ exclusion region is derived
at each mass point. These $Z'$ and $W'$ exclusion regions are shown in
Fig.~\ref{f:c}.  The branching ratio of $Z'$ or $W'$ to fermions
decreases as $\xi$ increases. This is opposite to the diboson decay
modes where branching ratios increase as $\xi$ increases.  Most $Z'$
or $W'$ search results~\cite{c:bk, c:wd0} report mass limits along the
$\xi = (M_{W}/M_{V})^{2}$ line and we have also done so for
comparison. However, the diboson decay modes and the fermionic decay
modes are sensitive to different parts of the gauge coupling strength
phase space, so searches for bosonic and fermionic decays of $Z'$ and
$W'$ are complementary to each other.  The $W'$ result shown in
Fig.~\ref{f:c} is significantly improved compared to the previous
result from CDF Run~I~\cite{c:wprime}. The $Z'$ result shown is the
first to set an exclusion region as a function of $\xi$ and mass.

In conclusion, we have searched for new particles decaying into a pair
of bosons in the electron, $\not\!\!\!E_{T}$, and two jets final state. In data
from an integrated luminosity of 2.9 fb$^{-1}$, no significant excess
over the standard model prediction is observed. Cross section limits
at 95\% C.L. and mass exclusion regions have been obtained for a
Randall-Sundrum graviton, $Z'$ and $W'$ bosons. The $W'$ exclusion
region in the $\xi-M_{W'}$ plane has been extended significantly
compared to the previous measurement. We have also presented the $Z'$
exclusion region in the $\xi-M_{Z'}$ plane for the first time. We set
the most stringent mass limits on $W'$ and $Z'$ bosons.


We thank the Fermilab staff and the technical staffs of the
participating institutions for their vital contributions. This work
was supported by the U.S. Department of Energy and National Science
Foundation; the Italian Istituto Nazionale di Fisica Nucleare; the
Ministry of Education, Culture, Sports, Science and Technology of
Japan; the Natural Sciences and Engineering Research Council of
Canada; the National Science Council of the Republic of China; the
Swiss National Science Foundation; the A.P. Sloan Foundation; the
Bundesministerium f\"ur Bildung und Forschung, Germany; the World
Class University Program, the National Research Foundation of Korea;
the Science and Technology Facilities Council and the Royal Society,
UK; the Institut National de Physique Nucleaire et Physique des
Particules/CNRS; the Russian Foundation for Basic Research; the
Ministerio de Ciencia e Innovaci\'{o}n, and Programa
Consolider-Ingenio 2010, Spain; the Slovak R\&D Agency; and the
Academy of Finland.




\begin{thebibliography}{99}
\bibitem{c:cdf_coord} D. Acosta {\it et al.}, Phys. Rev. D {\bf 71},
  032001 (2005).  CDF uses a cylindrical coordinate system in which
  $+z$ points along the direction of the proton beam, $r$ is the
  radius from the nominal beam line, and $\phi$ is the azimuthal
  angle.  The pseudorapidity is defined as $\eta=-\ln[\tan(\theta/2)]$,
  where $\theta$ is the polar angle measured from the $+z$
  axis. Transverse energy is defined as $E_{T} = E\cdot\sin\theta$, where
  $E$ is the measured calorimeter energy. Transverse momentum is
  defined as $p_T = p\cdot\sin\theta$, with $p$ being the track momentum.
\bibitem{c:iso} Electrons are isolated if in a surrounding cone of 0.4
  radius the $E_T$ deposited is no more than 1.1 times the electron
  $E_T$ in the EM calorimeters.
\end{thebibliography}
\end{document}